\documentclass[a4paper, 11pt]{article}
\date{December 2005}
\usepackage{epsfig}
  
\newcommand{\be}{\begin{equation}}
\newcommand{\ee}{\end{equation}}
\newcommand{\ba}{\begin{eqnarray}}
\newcommand{\ea}{\end{eqnarray}}
\newcommand{\bi}{\begin{itemize}}
\newcommand{\ei}{\end{itemize}}
\newcommand{\tr}{{\rm Tr\,}}
\newcommand{\re}{\mathop{\rm Re}}

\newcommand{\half}{{\textstyle\frac{1}{2}}}
\newcommand{\sigeff}{\sigma}
\newcommand{\pitf}{{\textstyle\frac{\pi}{24}}}
\newcommand{\N}{{\cal N}}
\newcommand{\rs}{r_{\rm s}}
\newcommand{\ceff}{c}

\newcommand{\nperp}{n_\perp}
\newcommand{\<}{\langle}
\renewcommand{\>}{\rangle}
\newcommand{\eq}{Eq.~}
\newcommand{\sm}{\small}

\newcommand{\fig}{Fig.~}
\newcommand{\tab}{Tab.~}
\newcommand{\la}{\label}

\begin{document}
\begin{titlepage}
DESY 06-109\\
 
\begin{centering}
\vfill

 \vspace*{2.0cm}
{\bf \Large Static forces in $d=2+1$ SU($N$) gauge theories}

\vspace{2.0cm}
{\bf Harvey~B.~Meyer}
\centerline{Deutsches Elektronen-Synchrotron DESY}
\centerline{Platanenallee 6}
\centerline{D-15738 Zeuthen}
\vspace{0.1cm}\\
\centerline{harvey.meyer@desy.de}

\vspace*{2.0cm}

\end{centering}
\centerline{\bf Abstract}
\vspace{0.1cm}
\noindent
Using a three-level algorithm 
we perform a high-precision lattice computation of 
the static force up to 1fm in the 2+1 dimensional SU(5) gauge theory.
Discretization errors and the continuum limit are discussed in detail.
By comparison with existing SU(2) and SU(3) data
it is found that $\sigma r_0^2=1.65-\pitf$ holds at an accuracy 
of $1\%$ for all $N\geq2$, where $r_0$ is the Sommer reference scale.
The effective central charge $\ceff(r)$ is obtained and
an intermediate distance $\rs$ is defined via the property
$\ceff(\rs)=\frac{\pi}{24}$. It separates in a natural way
the short-distance regime governed by perturbation theory 
from the long-distance regime described by an effective string theory.
The ratio $\rs/r_0$  decreases significantly 
from SU(2) to SU(3) to SU(5), where $\rs<r_0$. 
We give a preliminary estimate of its value in the large-$N$ limit.
The static force in the smallest representation of $\N$-ality 2,
which tends to the $k=2$ string tension as $r\to\infty$, is 
also computed up to 0.7fm. The deviation
from Casimir scaling is positive and grows from $0.1\%$ to $1\%$ 
in that range. 
\vfill
\end{titlepage}

\setcounter{footnote}{0}
\section{Introduction\la{sec:intro}}

One of the prime observables giving an insight into the 
workings of four-dimensional SU($N$)
gauge theories is the force $F(r)$ which two static quarks
separated by a distance $r$ exert on each other.
It may for instance serve to define a renormalized coupling constant,
and hence the knowledge of $F(r)$ at all distances
allows one to extract $\Lambda_{\rm \overline{MS}}^2$ in units
of the string tension $\sigma$,
establishing the connection between the regime accurately described
by perturbation theory and the regime thought to be described
by an effective string theory.
While in practice finite-volume schemes are better suited to overcome 
the hierarchy problem in connecting the very-short-distance regime to
the non-perturbative regime~\cite{ALPHA},
$F(r)$ is ideally suited to probe the 
string properties of the chromo-electric flux lines, since its functional form 
at asymptotic distances is a central prediction of 
effective string theories~\cite{polstrom,lw-beyond}.

Although the physically most important SU($N$) gauge theories
are undoubtedly defined in $d=4$ dimensions, in this paper 
we shall focus on the $d=3$ theories. Apart from being 
interesting in their own right (they exhibit a mass gap and linear
confinement, as demonstrated numerically in~\cite{teper2+1}), 
the $N=3$ case describes the magnetic sector
of QCD at asymptotically high temperatures and is thus relevant 
to the description of real-world physics 
(see~\cite{mikko} for a review on this aspect).

The theme of this paper is the $N$-dependence of static forces.
Due to super-renormalizability in $d=3$, 
the quantity analogous to $\sqrt{\sigma}/\Lambda_{\rm \overline{MS}}$
is $\sqrt{\sigma}/(g^2N)$, which has been calculated in~\cite{teper2+1}. 
Its $N$-dependence is well described by a constant, plus small
O$(1/N^2)$ corrections. This means that once two SU($N$) 
theories have been matched in the ultraviolet (as originally 
proposed by t'Hooft~\cite{hooft-largeN}), their far infrared behaviour
will agree too; and vice versa, these statements holding 
up to O$(1/N^2)$ corrections. The low-lying spectra of these theories,
compared in units of $\sqrt{\sigma}$,
are also very similar~\cite{teper2+1,regge2+1,thesis}.
The static force $F(r)$ gives us an independent way to compare
the theories at an adjustable distance scale.

The asymptotic approach to a constant force is of particular interest,
because the L\"uscher term~\cite{luscher81}, as well as the 
next term in the $(\sigma r^2)^{-1}$ expansion~\cite{lw-beyond},
are universal. Thus by comparing the static forces of different
SU($N$) gauge theories, one obtains a test of universality.
By the same token, the pre-asymptotic differences between
these functions of $r$  inform  us on the $N$-dependence of 
higher order string corrections,
as well as on the terms that vanish at long distance faster than any power of $1/r$.
Indeed it is widely believed~\cite{dalley,kuti_dublin} 
that the flux-tube also admits massive 
modes, in particular longitudinal compression modes.
How their effects are to be disentangled when only a finite number 
of terms of the asymptotic series are known is however not clear to us.

What can we expect about the $N$-dependence of the higher-order 
coefficients in the $1/r$ expansion of the static force? Since
the latter is related by open-closed string duality~\cite{lw-beyond} 
to the spectrum of torelons in the theory defined on a spatial $L\times\infty$ 
cylinder, and that the spectrum is expected to have a finite large-$N$ limit
with $1/N^2$ corrections at all $L$, one would expect (although it 
is not mathematically guaranteed) the 
coefficients of the $1/r$ series to have a finite large-$N$ limit with 
$1/N^2$ corrections.
Note that the perturbative series in $g^2N$, which is also 
asymptotic, does have this property, at 
least in the low orders where the coefficients are computed explicitly.

Powerful Monte-Carlo techniques were developed in~\cite{lw-algo} that 
triggered a very accurate determination of $F(r)$ in the range
$0.2{\rm fm}< r < 1.2{\rm fm}$ (throughout this paper 
the `fermi' is identified with $2r_0$, where $r_0$ is 
the Sommer reference scale~\cite{sommer}). Here
we extend these techniques to address the following physics issues:
\begin{itemize}
\item what is the $N$-dependence of $F(r)/\sigma$? A special case is 
the separation $r=r_0$, where
this amounts to studying the $N$-dependence of $r_0^2\sigma$;
\item since the $N$-dependence of $r_0^2\sigma$ turns out to be 
very weak, we then ask about the $N$-dependence of the effective
central charge $c(r)\equiv -\frac{1}{2}r^3\frac{dF(r)}{dr}$.
Any effective theory whose light degrees of freedom correspond to the 
transverse fluctuations of the flux-tube predicts 
$\lim_{r\to\infty} c(r) = \frac{\pi}{24}(d-2)$, independently
of $N$~\cite{lw-bosonic}. 
However the approach to this asymptotic value does have an 
$N$-dependence which turns out to be substantial;
\item what is the dependence of the static force on the color 
representation of the charges?
Many confinement models predict the static force 
in a representation ${\rm R}$
to be proportional to its quadratic Casimir 
$C_{\rm R}$~\cite{ambjorn3d,ambjorn4d,greensite,cornwall,simonov}.
This question has been addressed extensively 
(in $d=4$~\cite{deldar,bali,piccioni}, but also in 
$d=3$~\cite{poulis,krato,Stephenson:1999kh,Philipsen:1999wf});
our aim will be to quantify the small deviations from the 
`Casimir scaling' prediction.
\end{itemize}

Concerning this last point, 
we focus on the representations of two `quarks' 
symmetrized or antisymmetrized in color in the $d=3$ SU(5) theory.
These are particularly interesting, as the corresponding static forces
are known~\cite{teper-lucini} to be linearly confining with a string tension 
larger than in the fundamental representation. Since perturbation theory
predicts Casimir scaling of the static forces to hold at short distances~\cite{schroeder}, 
and that the string tensions have been found to be near-proportional to the 
smallest Casimir of the given $\N$-ality~\cite{teper-lucini,altes-thick}, 
it is \emph{a priori} plausible that Casimir 
scaling of the static force provides a good approximation at all separations~$r$
in the antisymmetric case.
The situation is thus different from studies in SU(2) or SU(3), where 
the static force in any representation asymptotically vanishes or 
is given by the fundamental string tension. The difference is due to the 
$Z(N)$ symmetry, which protects the $k=2$ string from screening once
$N\geq4$.

Note that $r_0^2\sigma$, $c(r)$ and $F_{\rm R}(r)/F_{\rm F}(r)$
(the index refers to the  representation, F being the fundamental one)
are all quantities which have a continuum limit -- unlike
for instance $r_0V(r)$, the static potential in units of $r_0^{-1}$.
Because we are after small effects (corrections of $c(r)$ from $\pi/24$,
deviation of $F_{\rm R}(r)/F_{\rm F}(r)$ from $C_{\rm R}/C_{\rm F}$),
the O($a^2$) discretization errors need to be at least estimated, or even better
the continuum extrapolation must be carried out.

Accurate SU(3) data was obtained a few years ago with a multi-level
algorithm in $d=3$ and $d=4$~\cite{lw-bosonic} 
(in the latter case also more recently in \cite{HariDass:2005we}), 
and SU(2) data is also available in $d=3$~\cite{pushan,caselle}.
Here we perform a $d=3$ SU(5) calculation that supplements the existing data.
We do so by generalizing the multi-level algorithm~\cite{lw-algo}
for the higher representations (see also~\cite{krato} for the adjoint case),
and introduce a further level of factorization~\cite{kari,martinh} of the Polyakov loop
correlator from which the static force is extracted.

The algorithmic and technical details are given in section~\ref{sec:simul}.
The new static force data for SU(5) is presented in section~\ref{sec:F5},
and the $N$-dependence of $F(r)$ 
discussed in~\ref{sec:r02sig}. Sections~\ref{sec:c5} and 
\ref{sec:cN} concern the effective central charge;
in the latter the SU(5) data is compared to the 
existing SU(2) and SU(3) data. Section~\ref{sec:FkF1} discusses the ratios of 
static forces and their deviations from Casimir scaling.
Our conclusions are gathered in section~\ref{sec:conclu}.
The appendix contains a discussion of discretization errors and of the continuum limit.

\section{Lattice simulations\la{sec:simul}}
In this section we describe the technical details of the computation.
The number of dimensions $d$ and the number of colors $N$ are still 
kept unspecified at this stage.

\subsection{Action and observables}

We use the standard plaquette action:
\ba
S&=&\frac{\beta}{N} \sum_{\rm plaq} \re\tr_{\rm F}\left\{1-U_p\right\}, \qquad
\beta = \frac{2N}{a^{4-d}g_{\rm o}^2}.
\la{eq:action}
\ea
where $U_p\in SU(N)$ is the ordered product of links around a plaquette. The quantity
$\tr_{\rm R}U$ is the trace of $U$ in the representation R; the subscript R=F corresponds to the 
fundamental representation. The size of the lattice is $L^{d-1}T$, and the boundary conditions for 
the link variables are periodic in all directions.

Our primary observable is the Polyakov loop correlator
\be
\<P_{\rm R}^*(x)P_{\rm R}(y)\> = \frac{1}{Z}\int D[U]~ P_{\rm R}^*(x)P_{\rm R}(y) e^{-S[U]},\qquad
D[U]=\prod_{x,\mu} dU_\mu(x) \la{eq:Pcor}
\ee
where $dU$ is the normalized invariant measure, $Z$ is such that $\<1\>=1$ and
\be
P_{\rm R}(x) = \tr_{\rm R}\{U_P(x)\},\qquad    U_P(x)=U_0(x)U_0(x+a\hat0)\dots U_0(x+(T-a)\hat0).
\la{eq:Pdef}
\ee
The trace may be taken in a general representation R of SU($N$). Since the link
variables belong to the fundamental representation, we relate $\tr_{\rm R}U$ to the trace
in the fundamental representation. Here are some examples for three irreducible representations:
\ba
\tr_{\rm A}\{U\} &=&  |\tr_{\rm F}\{U\}|^2 - 1  \la{eq:A}\\
\tr_{\rm 2a}\{U\} &=& \half\left((\tr_{\rm F}\{U\})^2 - \tr_{\rm F}\{U^2\}\right) \la{eq:2a} \\
\tr_{\rm 2s}\{U\} &=& \half\left((\tr_{\rm F}\{U\})^2 + \tr_{\rm F}\{U^2\}\right) .  \la{eq:2b}
\ea
The representation A is the adjoint representation, 2a and 2s are the representations of two quarks 
(anti)symmetrized in color. We have computed the Polyakov loop correlator for R=F, 2a and 2s.
Throughout this paper the index `1', or an index omitted altogether, 
stands for the fundamental representation F.

We now describe the algorithmic details.

\subsection{The multi-level algorithm\la{sec:algo}}

Rather than alternating full-lattice sweeps with measurements,
the algorithm we employ
proceeds in a multi-level scheme which exploits the manifest locality of the
action (\ref{eq:action}) to factorize the Polyakov loops correlators into several more local
functionals of the gauge field. 
The short distance fluctuations of the fields may then be averaged out independently
on each of these factors.

\subsubsection{Correlator in the fundamental representation}
The factorization which yields the largest gain is the one introduced by 
L\"uscher and Weisz~\cite{lw-algo}. In their algorithm, the two (untraced) 
Polyakov loops are sliced up into $n_0=T/\Delta_0$ segments by $n_0$ equidistant time-slices.
Recall the direct product of two matrices,
\be
(U\otimes V)_{\alpha\beta\gamma\delta} = U_{\alpha\beta} V_{\gamma\delta}.\la{eq:dp}
\ee
We first introduce the one-line operator
\be
{\bf T}^{\uparrow}_s(x) = {\cal T}~{\textstyle \prod_{j=0}^{\Delta_0/a-1}}~ U_0(x_{[s]}+aj\hat0).
\la{eq:1l}
\ee
It depends only on the temporal links between time-slice $s\Delta_0/a$ and $(s+1)\Delta_0/a$, 
time-slice $T/a$ being identified with time-slice 0. For a point $x$
with $x_0=0$, we defined $x_{[s]} = x + s\Delta_0\hat0$, $s=0,\dots,n_0-1$;
$x$ and $y$ can of course be taken to lie in the $x_0=0$ time-slice without loss
of generality. The symbol ${\cal T}$ means that the product is time-ordered.
Next we introduce a two-line operator:
\ba
{\bf T}^{\downarrow\uparrow}_s(x,y) &=& {\bf T}^{\uparrow}_s(x)^* \otimes {\bf T}^{\uparrow}_s(y).
\la{eq:2l}
\ea
Recall the product of two such direct-product matrices
\be
(U\otimes V) \cdot (U'\otimes V') = (UU')\otimes (VV'),
\ee
and the trace
\be
{\bf Tr}\{U\otimes V\} = \tr\{U\}~\tr\{V\}.
\ee
The Polyakov loop operator in the fundamental representation can then be written as 
\be
P_{\rm F}(x)^*P_{\rm F}(y) = {\bf Tr}\left\{ {\cal T}~{\textstyle\prod_{s=0}^{n_0-1}}~
{\bf T}^{\downarrow\uparrow}_s(x,y)\right\}.
\la{eq:Pf}
\ee

Because the contribution to the action of a particular link variable only depends on the local
staples, the Boltzmann weight $e^{-S[U]}$ also factorizes and one can write
\be
\< P_{\rm F}(x)^*P_{\rm F}(y)\> =  \<{\bf Tr}\left\{{\cal T}~{\textstyle\prod_{s=0}^{n_0-1}}~
                                            \<{\bf T}^{\downarrow\uparrow}_s(x,y)\>_s \right\}\>
\la{eq:Pfvev}
\ee
where $\<.\>_s$ denotes the average with the same Boltzmann weight but with fixed
spatial links in time-slices $s\Delta_0/a$ and $(s+1)\Delta_0/a$.
This formula is the basis of the algorithm introduced by L\"uscher and Weisz~\cite{lw-algo}.

We shall exploit an additional factorization level~\cite{kari,martinh}, 
by which $\<{\bf T}^{\downarrow\uparrow}_s(x,y)\>_s$ can be estimated
as follows. Single out the direction $\hat1$ and decompose the slab of volume $L^{d-1}\cdot\Delta_0$
into $n_1=L/\Delta_1$ smaller blocks of volume $\Delta_1\cdot L^{d-2}\cdot\Delta_0$. 
\fig\ref{fig:slablock} may help to visualize the situation. We then have
\be
\<{\bf T}^{\downarrow\uparrow}_s(x,y)\>_s = 
\< {\bf T}^{\uparrow}_s(x)\>_{sb_x}^*\otimes\<{\bf T}^{\uparrow}_s(y)\>_{sb_y}.
\la{eq:bvev}
\ee
The blocks labeled by the integers $b_x=[x_1/\Delta_1]$ and $b_y=[y_1/\Delta_1]$ 
must be different, with $x$ belonging to the former and 
$y$ belonging to the latter. The expectation value $\<.\>_{sb}$ is taken with 
fixed spatial links in time slices $s\Delta_0$ and $(s+1)\Delta_0$ and 
also fixed links in the set
\be
\left\{ U_\mu(z) ~|~ z_1=b_z\cdot \Delta_1 ~~{\rm and}~~ \mu\neq1\right\}.
\ee

\begin{figure}
\begin{center}
\psfig{file=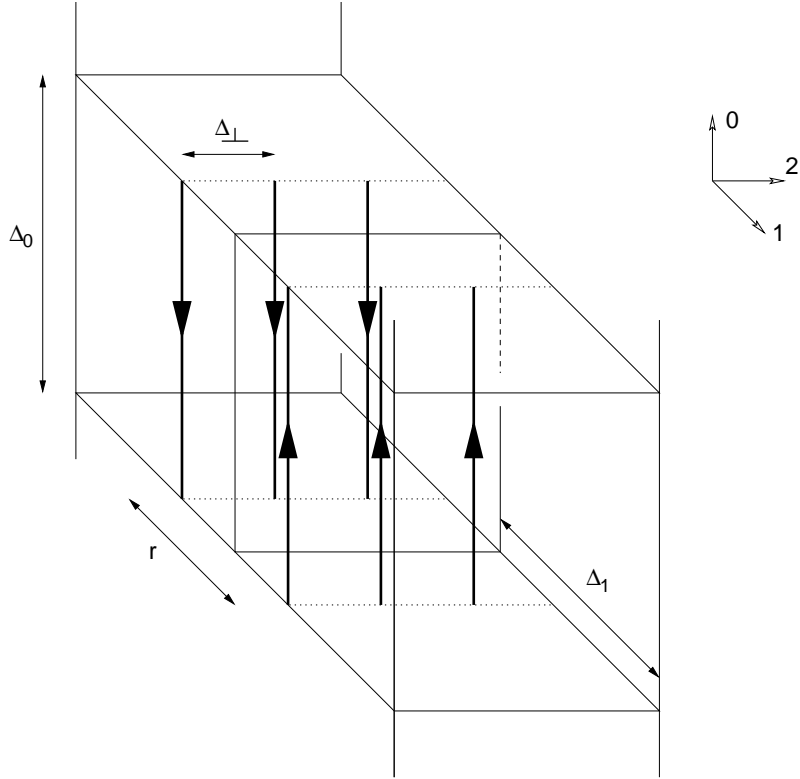,angle=0,width=10.5cm}
\end{center}
\caption{The geometry of slabs and blocks in the $d=3$ case. The thick lines symbolize
the Polyakov loops, and $\Delta_\perp n_\perp=L$.}
\la{fig:slablock}
\end{figure}

\subsubsection{Higher representations}

To incorporate the Polyakov loop correlator in higher representations in the multi-level
scheme described above, higher direct products must be computed. We generalize \eq\ref{eq:dp}
so that, for any two color tensors $U$ and $V$ of rank $n$  and $m$, 
(the color indices taking values 1 to $N$),
\be
(U\otimes V)_{\alpha_1,\dots,\alpha_n;\beta_1,\dots,\beta_m}
= U_{\alpha_1,\dots,\alpha_n} V_{\beta_1,\dots,\beta_m}.
\ee
With the definition
\be
{\bf T}^{\uparrow\uparrow}_s(x,y) = {\bf T}^{\uparrow}_s(x)  \otimes  {\bf T}^{\uparrow}_s(y),
\la{eq:2luu}
\ee
we can define the $\N$-ality 2 version of the two-line operator
\be
{\bf T}^{\downarrow\downarrow\uparrow\uparrow}_s(x,y) = 
{\bf T}^{\uparrow\uparrow}_s(x,x)^* \otimes {\bf T}^{\uparrow\uparrow}_s(y,y),
\la{eq:2ldduu}
\ee
which has $N^4\times N^4$ entries. The product of two such objects is defined by 
the property 
\ba
\left((U\otimes U')\otimes (U''\otimes U''')\right)\cdot\left( (V\otimes V')\otimes (V''\otimes V''')\right)   
= (UV\otimes U'V') \otimes (U''V''\otimes U''' V''').\nonumber
\ea
and multilinearity.

The observables needed to compute the correlator in the 2a and 2s representations may now be expressed as
\ba
\la{eq:P2P2}
P_{\rm F}^2(x)^*P_{\rm F}^2(y) &=& {\bf Tr}_{\rm 2i}
\left\{ {\cal T}~{\textstyle\prod_{s=0}^{n_0-1}}~{\bf T}^{\downarrow\downarrow\uparrow\uparrow}_s(x,y)\right\},\\
\la{eq:U2U2}
\tr\{U_P(x)^2\}^*\tr \{U_P(y)^2\} &=& {\bf Tr}_{\rm 2ii}  
\left\{ {\cal T}~{\textstyle\prod_{s=0}^{n_0-1}}~{\bf T}^{\downarrow\downarrow\uparrow\uparrow}_s(x,y)\right\}, \\
\la{eq:P2U2}
P_{\rm F}^2(x)^*\tr \{U_P(y)^2\} &=& {\bf Tr}_{\rm 2iii}
\left\{ {\cal T}~{\textstyle\prod_{s=0}^{n_0-1}}~{\bf T}^{\downarrow\downarrow\uparrow\uparrow}_s(x,y)\right\},
\ea
where the trace operations are defined such that
\ba
{\bf Tr}_{\rm 2i} \{(U\otimes U')\otimes(V\otimes V') \} &=& \tr\{U\}~\tr\{U'\}~\tr\{V\}~\tr\{V'\}, \\
{\bf Tr}_{\rm 2ii}\{ (U\otimes U')\otimes(V\otimes V') \} &=& \tr\{UU'\}~\tr\{VV'\},  \\
{\bf Tr}_{\rm 2iii}\{ (U\otimes U')\otimes(V\otimes V') \} &=& \tr\{U\}~\tr\{U'\}~\tr\{VV'\}.
\ea

For the adjoint representation, one needs the definition
\be
{\bf T}^{\uparrow\downarrow\downarrow\uparrow}_s(x,y) = 
{\bf T}^{\downarrow\uparrow}_s(x,x)^*    \otimes {\bf T}^{\downarrow\uparrow}_s(y,y).
\la{eq:2luddu}
\ee
Then
\ba
\la{eq:Pa1}
|P_{\rm F}(x) P_{\rm F}(y)|^2 &=& {\bf Tr}_{\rm 2i} 
\left\{ {\cal T}~{\textstyle\prod_{s=0}^{n_0-1}}~{\bf T}^{\uparrow\downarrow\downarrow\uparrow}_s(x,y)\right\}\\
|P_{\rm F}(x)|^2 &=&  {\bf Tr}\left\{ {\cal T}~{\textstyle\prod_{s=0}^{n_0-1}}
{\bf T}^{\downarrow\uparrow}_s(x,x)\right\} 
\la{eq:Pa2}
\ea

As in going from \eq\ref{eq:Pf} to \eq\ref{eq:Pfvev}, 
the expectation values of the correlators can now be written in a form suitable 
for the implementation of the multi-level algorithm. In the expectation value of the 
Polyakov loop correlators one may take the average value $\<.\>_s$ inside a slab of each 
two-line operator appearing
in \eq\ref{eq:P2P2}---\ref{eq:P2U2}, \ref{eq:Pa1}---\ref{eq:Pa2}.
To exploit the factorization into blocks along the direction $\hat1$, 
we evaluate the slab-averages as in \eq\ref{eq:bvev} by taking the block average values of
the line operators on the right-hand side of \eq\ref{eq:2ldduu} and \eq\ref{eq:2luddu}.

It should be clear that the Polyakov loop correlator can be obtained in a factorized form 
for any desired representation. Also, the formulae presented hold for any number of dimensions
$d\geq2$. We now turn to some practical considerations concerning
the implementation of the algorithm.
\subsubsection{Algorithm implementation and efficiency}

\begin{table}
\begin{center}
\fbox{\begin{minipage}[l]{13.5cm}
\small{\begin{enumerate}
\item Produce an independent configuration by doing a sufficient number of 
      hybrid over-relaxation sweeps through the whole lattice;
\item for all slabs, in increasing time order: 
\item $\qquad$ repeat the following procedure $m_s$ times:
\item $\qquad\qquad$ update the slab and then, for all blocks: 
\item $\qquad\qquad\qquad$ do $m_b$ measurements of the needed line operators  

      $\qquad\qquad\qquad$ inside the block separated by an update within the block
\item $\qquad\qquad$ compute the two-line operators and add them to their slab averages;
\item $\qquad$ multiply the product of previous two-link operators by the new one;
\item compute the traces.
\end{enumerate}}
\end{minipage}}
\end{center}
\caption{Outline of the three-level algorithm.}
\la{tab:algo}
\end{table}

The three-level algorithm described above that we applied to $d=3$ SU(5) simulations
is in essence the L\"uscher-Weisz algorithm, where however 
the slab-averages of the two-line operators are computed in a factorized way. 
Due to the  factorization along the direction~$\hat1$, we measure the Polyakov loop correlator 
in that direction only. The additional factorization breaks translational invariance
and the symmetry between directions~$\hat1$ and~$\hat2$ and thus leads to a loss of statistics;
on the other hand the short-distance fluctuations of the one-line operators 
can be averaged out separately. Test studies~\cite{inprep} in the $d=3$ SU(2) theory
show that the three-level algorithm improves on 
the original L\"uscher-Weisz algorithm for $r\geq0.9$fm, $a=0.08$fm,
the latter being superior at shorter distances. This test concerns the static force
in the fundamental representation, but in the higher representations, 
and also at smaller lattice spacing,
we expect the additional factorization to pay off already at shorter distances. 
Since the correlation is obtained at fewer points, but each of them is evaluated more 
accurately, the memory requirement is lowered. 

\begin{table}
\begin{center}
\begin{tabular}{c@{\quad}c@{\quad}c@{\quad}c@{\quad}c@{\quad}c@{\quad}c@{\quad}c@{\quad}l}
\hline\hline
        &$\beta$    & $T/a$ & $L/a$ & $n_0$ & $n_1$ & $\nperp$ & $n_{\rm meas}$ &   $r_0/a$\\
\hline
$A_{20}$  & 38       &  20   &  22  & 4   & 2 & 11 & 2536   & 4.03879(39) \\
$A_{30}$  & 38       &  30   &  22  & 6   & 2 & 11 &  2275   &  4.03943(35) \\
$B$       & 44       &  25   &  26  & 5   & 2 & 13 &  1731   &4.82879(58) \\
$C$       & 54       &  30   &  30  & 5   & 3 & 10 & 1647 &  6.1323(11) \\
\hline\hline
\end{tabular}
\end{center}
\caption{The $d=3$ SU(5) simulation details.}
\la{tab:simdat}
\end{table}

The algorithm proceeds as sketched in \tab\ref{tab:algo} (the scope of loops are 
given by the indentation of the lines).
The simulation parameters are given in \tab\ref{tab:simdat}.
The number $n_0$ of slabs and $n_1$ of blocks per slab can be found there.
Because of memory limitations, we only computed the line-operators 
every second or third point in the $\hat2$ direction. The number $\nperp$ in \tab\ref{tab:simdat}
is the number of transverse coordinates at which the line-operators were computed. 
We measured the correlation from distance $2a$ to $10a$ in all simulations; hence $n_r=9$.
And in all simulations $m_s=20$ and $m_b=60$ (see \tab\ref{tab:algo} for the meaning
of these parameters). We did not use the multihit technique~\cite{multihit},
because for the higher representations
it increases the number of two-line operator multiplications, which are expensive at large $N$.
Finally, the total number of update sweeps between measurements is $60$. They are grouped 
in compound sweeps consisting of one heat-bath followed by three over-relaxation 
sweeps~\cite{fabhaan,kenpen,adler-or}.
This proved sufficient to decorrelate almost entirely successive measurements of the 
observables and still represents a negligible overhead with respect to the measurements. 
To update the slabs we used three compound sweeps, 
and only one between line-operator measurements inside the blocks.

The program goes from slab to slab in a time-ordered way.
It is in order to save memory that each measurement of the two-line operator slab-averages
is followed by their multiplication with the product of the preceding slab-averages. 
In this way, the program only needs enough memory for two fields of the size
\[ n_1 \times \nperp \times n_r \times {\rm size}({\bf T}).\]
The two-line operator (\ref{eq:2ldduu}) contains $N^8$ complex numbers.
Hence we reach the requirement $\sim4\cdot10^8$ 32-bit floating numbers for simulation C.

For a given separation $r$, we correlate the line operators as far as possible from the boundaries. 
That is, for the correlator at a separation of $2na$, we correlate
the line operators computed at $x_1=\Delta_1-na$ with those computed at $x_1=\Delta_1+na$ 
(and all equivalent points
by translation of distance $\Delta_1$ in the direction $\hat1$). For a separation of $(2n+1)a$,
we average the correlation obtained at  $(\Delta_1-na,\Delta_1+(n+1)a)$ with 
that obtained at $(\Delta_1-(n+1)a,\Delta_1+na)$.

We introduce the notation
\be
\Gamma_{\rm R}(r) = \<P_{\rm R}^*(x)P_{\rm R}(x+r \hat 1)\>
\ee
for the on-axis Polyakov loop correlator.
\tab\ref{tab:PP44} gives these correlators in the fundamental, as well as
in the $\N$-ality 2 representations 2a and 2s in simulation B. 
It turns out that these measurements in the different representations
are quite strongly correlated at short distances
(and essentially uncorrelated beyond $r=r_0$), so that
it will be advantageous to take the ratio of static forces measured in the 
same simulation. With the tuning of the algorithm used, 
we are able to maintain a signal to noise ratio below $1\%$
beyond 1fm in the fundamental representation and beyond 0.7fm in the 2a representation.
The performance (and memory requirements) of the algorithm 
behave favourably  if  the time extent is increased, 
as a comparison of the data from simulations $A_{20}$ and $A_{30}$ will reveal.

To give an idea of the computing effort involved, one full measurement in 
simulation $B$ took 1.7 hour on a single 2.4GHz Opteron with 4GB of memory,
while the 60 full-lattice sweeps between each of these full measurements
took 61 seconds.

The statistical errors are estimated with the Gamma method~\cite{uwerr}.
Simulation $C$ was run with 8 replica (= independent simulations)
and the others with 4 replica; we checked in all cases that the results are 
consistent across replica. We also found that using the jacknife method
gave error estimates in agreement with the Gamma method, 
which is hardly surprising since the successive measurements were very 
decorrelated in these runs.

\section{The static force in the fundamental representation\label{sec:FF}}
In this section we present our data on the static force. It is found 
that the quantity $r_0^2\sigma$ has very small discretization errors. 
By comparison with the SU(3) data of L\"uscher and Weisz~\cite{lw-bosonic}
and  the SU(2) data of Majumdar~\cite{pushan} (see also~\cite{caselle}), 
we find that it is also 
very weakly dependent on the number of colors $N$. We then compare the 
$N=2,3,5$ results for the effective central charge at fixed value of $r_0$.

\subsection{The SU(5) static force\label{sec:F5}}
The static force is defined at finite lattice spacing by 
\be
F_{\rm R}(\bar r) = \textstyle{\frac{1}{a}}[V_{\rm R}(r+a)-V_{\rm R}(r)],
\qquad V_{\rm R}(r) = -\frac{1}{T} \log\Gamma_{\rm R}(r).
\ee
We reuse the tree-level improved argument $\bar r=r+a/2+{\rm O}(a^2)$ 
of the function $F_{\rm R}$ given in~\cite{lw-bosonic}.
The fundamental static force obtained in simulation B can be found in \tab\ref{tab:Fr}.
The reference scale $r_0/a$ given in \tab\ref{tab:simdat} 
is obtained in the standard manner~\cite{sommer}, by linearly 
interpolating $(r/a)^2$ as a function of $r^2F(r)$ to the abscissa 1.65. 

To a first approximation, the force is constant beyond 0.75fm, signalling linear confinement.
To investigate this in more detail, we find it useful to define the quantity
\be
\sigeff(r) = F(r) - \frac{\pi}{24r^2}. \la{eq:sigeff}
\ee
If the L\"uscher term does provide the leading asymptotic correction
to the linear potential~\cite{luscher81}, then this quantity deserves the name of 
effective string tension and the notation is appropriate.
Note that by definition,
\be
r_0^2\sigeff(r_0) = 1.65-\pitf = 1.519100\dots
\ee
at all lattice spacings. In particular, discretization errors on 
$r_0^2\sigeff(r)$ are automatically reduced around separation $r=r_0$.

The function $r_0^2\sigeff(r)$ is plotted in \fig\ref{fig:sigeff} for all data sets.
The practical advantage of considering this function is that it is very flat beyond 0.4fm 
(note the scale of the vertical axis); in this way, the data can be visualized
in more detail.
We firstly remark that the data $A_{20}$ and $A_{30}$ agree within error bars at 
all separations; the common abscissa of these data points have been split symmetrically 
for better visibility.
This indicates that the time extent $T\approx 5r_0$ is sufficient 
to `filter out' the ground state of the transfer matrix in the presence of 
the two Polyakov loops~\cite{lw-bosonic} separated by $r\leq 2r_0$; 
the statement holds of course
at the precision to which we have been able to compute the static force.

Secondly, the scaling violations of $r_0^2\sigeff(r)$ are at the few
permille level (or less) around $r=1.4r_0$.
To carry out a continuum extrapolation, we need to interpolate 
$r_0^2F(xr_0)$ to a few fixed values of $x$. This is done
by a linear interpolation  in $1/x^2$
(in \cite{lw-bosonic} a three-point interpolation in $1/x^2$ was used). 
The result is given for each simulation in \tab\ref{tab:Finterpol}.
It turns out that extrapolating these quantities to the continuum with 
the expected ${\rm O}(a^2)$ corrections~\cite{sommer-necco} 
yields poor $\chi^2$ values (see \fig\ref{fig:conti}). 
Therefore we extrapolate the data from the 
two smaller lattice spacings linearly in $(a/r_0)^2$, and use a 
quadratic fit to all three lattice spacings to estimate the 
uncertainty associated  with the linear fit; 
we refer the reader to the appendix for more details.
At least one additional simulation at a lattice spacing of $a\approx0.1r_0$ 
is required to take the continuum limit in a way that does not blow up the 
final uncertainty.

Finally, we did not investigate finite-volume effects, but since 
$2.45{\rm fm}<L<2.72{\rm fm}$ in these simulations,
we do not expect them to affect our conclusions.

\subsection{The quantity $\sigma r_0^2$ for all $N$\la{sec:r02sig}}

\begin{figure}
\begin{center}
\psfig{file=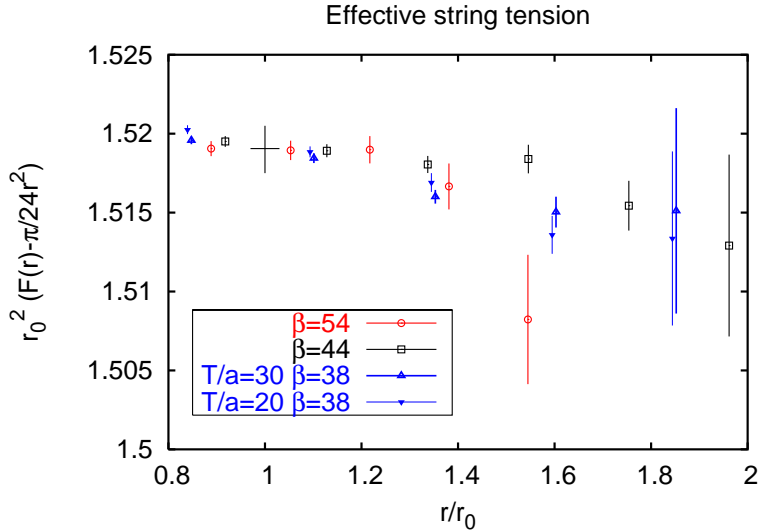,angle=0,width=10.5cm}
\end{center}
\caption{The SU(5) effective string tension in units of $r_0^{-2}$.
By interpolation all data go through the cross (by definition of $r_0$).}
\la{fig:sigeff}
\end{figure}

In comparing SU($N$) gauge theories, perhaps the most
natural way to set the common energy scale is by equating 
their (fundamental) string tensions $\sigma$. This quantity 
is however more difficult to compute with high precision
than $r_0$, because it involves an extrapolation to infinite string length~\cite{sommer}.
For that reason, we are now going to show that the dimensionless quantity 
$r_0^2\sigma$ is $N$-independent to a precision better than $1\%$.
As a consequence, comparing SU($N$) gauge theories
at fixed value of $r_0$ is essentially equivalent to fixing the value of $\sigma$.

We can extract the effective string tension at 1fm straightforwardly.
A direct comparison of our data with the SU(3) data of L\"uscher and Weisz 
at the same lattice spacing $a=0.207r_0$ gives
\ba
r_0^2\sigeff(2r_0) &=& 1.5124(65), \qquad {\rm SU}(5), ~\beta=44, \\
r_0^2\sigeff(2r_0) &=& 1.51776(24), \qquad {\rm SU}(3), ~\beta=15. 
\ea
Interestingly, the string tension has also been computed
in the closed string channel by Teper~\cite{teper2+1} at these values of $\beta$. Using
the values of $r_0$ quoted above, we have
\ba
{\rm SU}(5): && r_0^2~\sigma_{\rm closed} = 1.521(8) \\
{\rm SU}(3): && r_0^2~\sigma_{\rm closed} = 1.520(14).
\ea
The string tension quoted by Teper is also an effective string tension
(as defined by $m(L)=\sigeff(L)L-{\textstyle\frac{\pi}{6L}}$),
but at a closed string length of $3.3r_0$, where higher order string corrections
are presumably much reduced. In the SU(3) case an additional calculation of his
at string length of $5.0r_0$ yielded the same string tension (and accuracy).
We conclude that the closed string tension extracted from torelon masses
agrees with the open string tension extracted from the static force at 1fm
at a precision of $1\%$.

Majumdar~\cite{pushan} reports almost identical 
values of $r_0^2\sigma$ in the SU(2) theory at lattice spacings $a\leq 0.16r_0$.
In particular, a comparison between SU(2) and SU(3) closer to the continuum yields
\ba
r_0^2\sigeff(2r_0) &=& 1.523(2)\quad\qquad {\rm SU}(2), ~\beta=7.5,~r_0=6.29a \\
r_0^2\sigeff(2r_0) &=& 1.51999(34)\qquad {\rm SU}(3), ~\beta=20,~r_0=6.71a. 
\ea

Cutoff effects on $r_0^2\sigma$ are thus seen to lie well below 
the percent level. Assuming that no strong $N$-dependence appears
beyond $N=5$ (the contrary would be very surprising in view of the results obtained
in~\cite{teper2+1}), we reach the following remarkable conclusion:
\be
r_0^2\sigma = 1.52\cdot(1\pm1\%)\qquad \qquad\forall N\geq2,\quad a\leq0.21r_0.
\la{eq:r02sig}
\ee
In particular, 
\be
\sigma = \sigeff(r_0)\cdot(1\pm1\%)\qquad \qquad\forall N\geq2
\ee
holds in the continuum limit (see also \fig\ref{fig:conti}). 

\subsection{The effective central charge\la{sec:c5}}
\begin{figure}
\begin{center}
\psfig{file=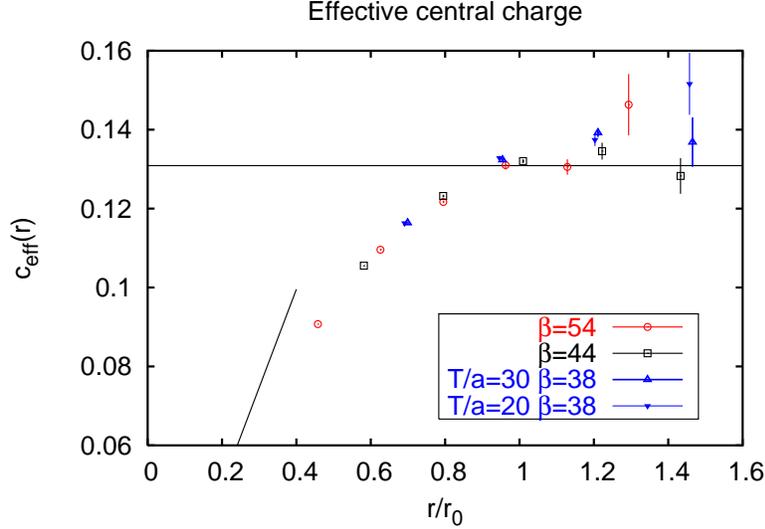,angle=0,width=10.5cm}
\end{center}
\caption{The SU(5) effective central charge at several lattice spacings.
The perturbative prediction is shown at short distance, and the horizontal line
marks the value $\pitf$.}
\la{fig:ceff5}
\end{figure}

The effective central charge is defined by 
\be
c(\tilde r) = -{\textstyle\frac{1}{2}}\tilde r^3 [V(r-a)-2V(r)+V(r+a)]/a^2,
\ee
where $\tilde r=r+{\rm O}(a^2)$ is defined in~\cite{lw-bosonic}.
\fig\ref{fig:ceff5} shows the computed effective central charge
$c(r)$ in the range of distances 0.25 to 0.75fm. A precocious convergence 
to values within $5\%$ of $\pitf$ is observed.

By eye, the data at $\beta=44$ and $\beta=54$ practically fall on top of each other 
up to 0.6fm, indicating that discretization errors are reasonably small. Beyond
that point, the $\beta=54$ data's accuracy no longer 
allows for a useful comparison. The data points obtained at
the coarsest lattice spacing systematically lie slightly above 
the other points. This effect was already seen in SU(2) and SU(3)
calculations. Also, the data sets $A_{20}$ and $A_{30}$ are well compatible,
indicating that the time extent was chosen long enough 
for the present statistical accuracy.

The expected behaviour at short distances, as given by two-loop perturbation
theory~\cite{schroeder}, is shown on the figure, as well as the universal value $\pitf$
expected at long distances. 
The continuum value of $g^2r_0$, which determines the slope of the short-distance
prediction, was estimated by extrapolating the value of $g_{\rm o}^2r_0$ of 
simulations $B$ and $C$ linearly in $g_{\rm o}^2a$. Using the continuum value
of $\sqrt{\sigma}/g^2$ of~\cite{teper2+1} and \eq\ref{eq:r02sig} yields a slope 
smaller by $2\%$; the difference would hardly be noticeable on \fig\ref{fig:ceff5}.

In view of taking the continuum limit, we interpolate
$c(xr_0)$ to a handful of values of $x$. We use a three-point polynomial interpolation 
in $1/x$ (following~\cite{lw-bosonic}); the result is given in \tab\ref{tab:cinterpol}. 
The continuum limit is described in the appendix, where also an alternative definition of 
the effective central charge is proposed and shown to lead to the same 
continuum results (\fig\ref{fig:conti}).

\subsection{The effective central charge for all $N$\la{sec:cN}}
\begin{figure}
\begin{center}
\psfig{file=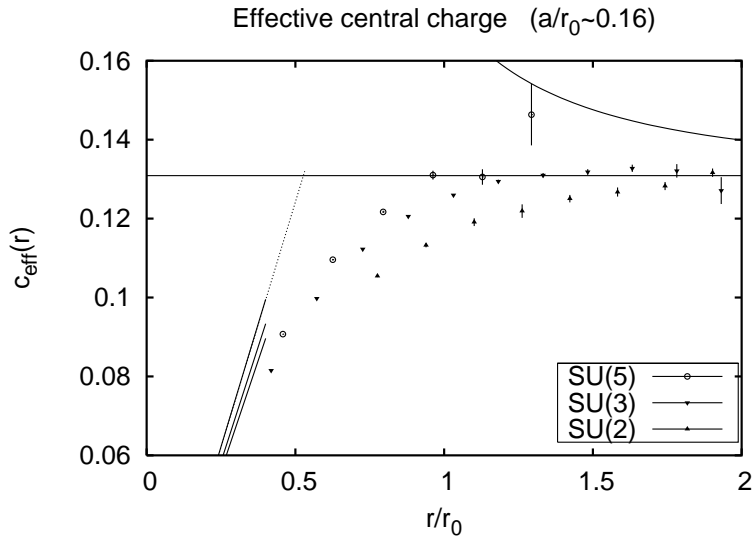,angle=0,width=10.5cm}
\end{center}
\caption{The effective central charge for several gauge groups.
The SU(3) data is taken from~\cite{lw-bosonic}, the SU(2) data
from~\cite{pushan-pc}.
The perturbative prediction is shown at short distance (the slope increases
slightly with $N$) and the Nambu-Goto prediction at long distances, 
while the horizontal line marks the value $\pitf$.} 
\la{fig:ceff}
\end{figure}

It is now interesting to compare the effective central charge curves
obtained for $N=2,3$ and 5. 
If we attempted to compare the data in the continuum limit,
the continuum extrapolation would lead to the dominant source
of uncertainty, because only three lattice spacings are
available at each $N$. Also, the data at the coarsest lattice spacing
in the SU(3) and SU(5) data sets are not (always) consistent 
with O($a^2$) corrections. For that reason we prefer to compare
the data sets at the smallest common lattice spacing. This corresponds
to $\beta=7.5,~20$ and 54 for $N=2,3$ and 5 respectively.
The values of $r_0/a$, 6.29, 6.71 and 6.13 respectively, indeed are
matched at the $10\%$ level.

The comparison is illustrated on \fig\ref{fig:ceff}. At short distances,
the curves are expected to differ by O($1/N^2$) from perturbation theory
and the scaling of $g^2Nr_0$. This is indicated by the straight solid lines
which stop short at $0.4r_0$ around where perturbation theory ceases 
to be accurate. Remarkably, all three curves seem to approach the value of 
$\pitf$ at larger distances. This fact constitutes
strong evidence for the universality of the string correction
in $d=3$ SU($N$) gauge theories.
On the other hand, the approach to the asymptotic value is surprisingly precocious.
The numerical closeness of $c(r)$ to
$\pitf$ is probably deceptive, in so far that it suggests that any 
higher order string corrections are already very small below 1fm.
The leading correction term is O($1/\sigma r^2$), and its coefficient
is universal and \emph{positive} in the effective theory describing 
the massless transverse fluctuations of the flux-tube~\cite{lw-bosonic}.
The Nambu-Goto string in a non-critical number of dimensions 
corresponds to a special case in this class of theories;  
it provides a consistent analytic prediction for the $r\times T$
Polyakov loop correlator to any finite order in $1/r$ or  $1/T$~\cite{lw-bosonic}.
It predicts an effective central charge~\cite{arvis} of the form
\be
c^{\rm (NG)}(r) = \frac{\pi}{24}~
\left(1-{\textstyle\frac{\pi}{12\sigma r^2}}\right)^{-3/2},
\la{eq:c_ng}
\ee
shown for illustration in the upper right corner of  \fig\ref{fig:ceff}.
It suggests that the effective string theory becomes accurate at significantly 
larger distances. To summarize, since $c(r)<\pitf$ at short distances; since 
the $1/r$ series for $c(r)$ at large distances is 
almost certainly asymptotic; and since the O($1/\sigma r^2$) term is positive, 
this term improves on the leading order prediction at the earliest when
the effective central charge computed in the gauge theory rises above
the value $\pitf$. 

The preceding arguments thus motivate the definition 
of the distance $\rs$ where the effective central charge crosses $\pitf$:
\be
c(\rs) \equiv \pitf   \la{eq:def r_s}.
\ee
This distance is a natural separation scale between the perturbative regime
and the string regime. 
The ratio $\rs/r_0$ is expected to have a continuum limit
with  O($a^2$) discretization errors, and at finite lattice spacing an 
interpolation formula must be prescribed to define it. In general we propose to use 
a three-point polynomial interpolation in $r$ using the three nearest data points,
because $c(r)$ has quite some curvature around $\rs$.

\begin{figure}
\begin{center}
\psfig{file=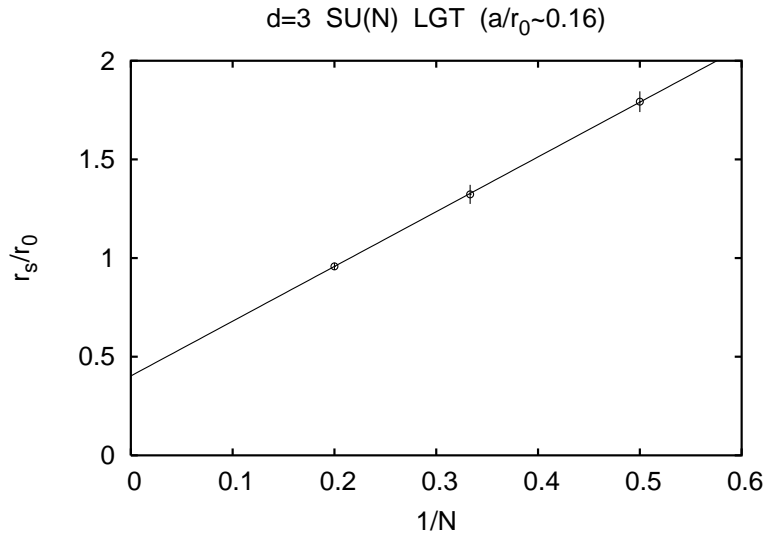,angle=0,width=10.5cm}
\end{center}
\caption{The dependence of $\rs/r_0$ on the number of colors $N$.
The SU(3) data is taken from~\cite{lw-bosonic}, the SU(2) data
from~\cite{pushan-pc}.}
\la{fig:rsr0allN}
\end{figure}

\fig\ref{fig:ceff} shows quite strikingly that as the number of colors 
is increased, the curve $c(r)$ interpolates over an even shorter
distance range between the perturbative behaviour and the value $\pitf$.
We therefore study the $N$-dependence of $\rs/r_0$. To do so we shall presently use 
a simple linear interpolation between the two nearest points to the crossing, because 
the SU(2) and SU(5) data rapidly become less accurate beyond the crossing.
Luckily in all three cases there is a point lying almost exactly at $\pitf$, which
makes the details of the interpolation a negligible source of uncertainty.
In order to have the same discretization scheme for all data sets, 
we replaced $r$ by $\tilde r$ in the SU(2) values of the effective central 
charge~\cite{pushan,pushan-pc}. We thus find:
\be
\frac{\rs}{r_0} = \left\{\begin{array}{l@{\quad}l}
1.79(5) & N=2 \\
1.32(5) & N=3 \\
0.96(2) & N=5 \\
 \end{array}\right. \la{eq:rsr0}
\ee
The smaller uncertainty at SU(5) reflects the fact that $c(r)$ cuts
the horizontal $\pitf$ line at a shorter distance and 
with a larger slope than for SU(2) and SU(3).
At $N=5,~\beta=44$ we obtain $\rs/r_0=0.977(11)$, in agreement with (\ref{eq:rsr0}).
The trend that $\rs/r_0$ strongly decreases with the number of colors is quite striking
and constitutes a rare example of a physical quantity in 
SU($N$) gauge theories which has a strong $N$-dependence.
This dependence is shown on \fig\ref{fig:rsr0allN}.
The data is well compatible with a linear function in~$1/N$:
\be
\frac{\rs}{r_0} = 0.402(45)  +  2.80(20)\frac{1}{N}\qquad
(\chi^2/{\rm d.o.f} \simeq 0.01).
\ee
Could it be that $\rs$ becomes as small as 0.2fm in the large-$N$ limit?
An SU(8) calculation would provide a useful clue to answer the question.
In the mean time, we can compare the value to two other distance scales
which are natural in this context.
The first comes from the Arvis formula, \eq\ref{eq:c_ng}. 
The latter becomes singular at a distance $r_a$ given by
\be
\frac{r_a}{r_0} = \sqrt{\frac{\pi}{12\sigma}} ~\frac{1}{r_0} = 0.415(4).
\ee
where we have used \eq\ref{eq:r02sig}.
It is intriguing that the naively obtained value $\rs(N=\infty)$ is compatible
with $r_a$. Secondly, the dotted line on \fig\ref{fig:ceff}
is the continuation of the curve corresponding to the perturbative two-loop result~\cite{schroeder}
in the large-$N$ limit 
(using $\lim_{N\to\infty}\sqrt{\sigma}/g^2N=0.1975(10)$ from~\cite{teper2+1} and \eq\ref{eq:r02sig}).
It cuts $\pitf$ at distance $r_{\rm pt}$ given by
\be
\frac{r_{\rm pt}}{r_0}=0.527(4).
\ee

The  rise of $c(r)$ up to and beyond $\pitf$, 
which  becomes  more rapid as $N$ is increased, could be a sign 
that the asymptotic series in $(\sigma r^2)^{-1}$ provided by the effective string
theory converges better at a given value of $\sqrt{\sigma} r$ if the number of colors
is increased. This is not unconceivable, 
because certain effects such as the radiation of glueballs
by a highly excited string, which exist in the gauge theory but 
are neglected in the effective string theory, are $1/N^2$ suppressed.

\section{What is the size of Casimir scaling violations?\label{sec:FkF1}}

\begin{figure}
\begin{center}
\psfig{file=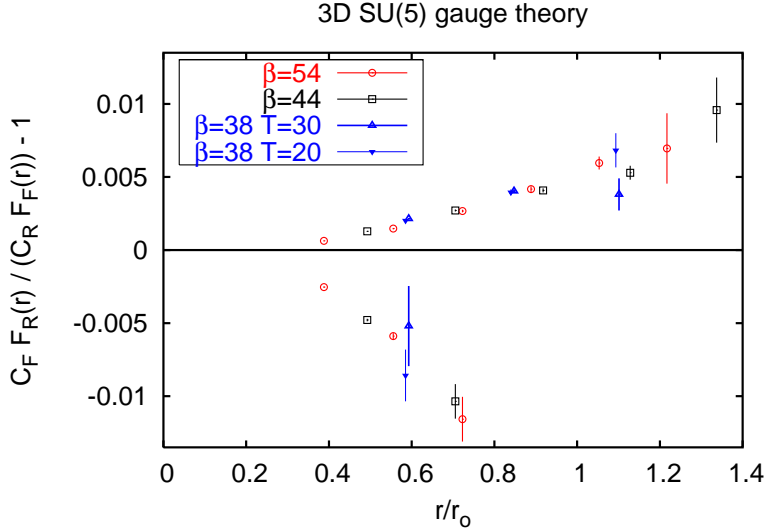,angle=0,width=10.5cm}
\end{center}
\caption{Relative deviation of the force ratios $F_{\rm R}/F_{\rm F}$ from Casimir scaling
in the SU(5) gauge theory. The points above and below
zero correspond to the representations R=2a and R=2s respectively.}
\la{fig:ratio}
\end{figure}

We now address this question in the case of the two smallest irreducible representations
of SU(5) of $\N$-ality 2. These are obtained by taking the direct product of 
two fundamental charges and forming the (anti)symmetric linear combinations, as 
in \eq\ref{eq:2a} and \ref{eq:2b}.
We discuss the ratio of these forces to the force in the fundamental representation.
Note that, unlike the ratio of static potentials, this quantity has a continuum limit.

The Casimir scaling prediction is, in a range of distances $r_{\rm min}< r <r_{\rm max}$, 
\ba
\frac{F_{\rm 2a}(r)}{F_1(r)} &\stackrel{\rm CS}{=}& \frac{C_{\rm 2a}}{C_1} =
\left.\frac{2(N-2)}{N-1}\right|_{N=5} = \frac{3}{2}.\\
\frac{F_{\rm 2s}(r)}{F_1(r)} &\stackrel{\rm CS}{=} & \frac{C_{\rm 2s}}{C_1} =
\left.\frac{2(N+2)}{N+1}\right|_{N=5} = \frac{7}{3}.
\ea
Perturbation theory predicts~\cite{schroeder}
\be
F_{\rm R}(r) = \frac{C_{\rm R} g^2}{2\pi r} 
\left[ 1 + \textstyle{\frac{7}{32}} g^2Nr + O((g^2Nr)^2)\right].
\ee
In particular, Casimir scaling of static forces holds at least to order $g^4$.

The result for the ratio of static forces obtained in simulation $B$ is given 
in \tab\ref{tab:Fr}.
\fig\ref{fig:ratio} shows the Monte-Carlo data on the relative deviations
of $F_{\rm 2a}/F_1$ from Casimir scaling
in the SU(5) lattice gauge theory. The window in $r$ is 0.2---0.7fm. Notice the scale of the 
vertical axis: the (2a) data points indicate a positive relative deviation which grows from about 0.001 to 0.01
in that range of distances. The (2s) data on the other hand shows a stronger, negative deviation.
This is expected, since the static charges in this representation can be screened to (2a).
The force between charges in the adjoint representation, which can be screened completely,
also exhibits this behaviour~\cite{krato}. A positive deviation is thus rather special. 
It will remain positive at larger distances if the $k=2$ string tension is
larger than $C_{\rm 2a}/C_1\cdot \sigma$, or if it is given exactly
by that expression and its central charge by $\pitf$.

Beyond $r_0$, we are perhaps seeing the effects of finite time extent: there is a 1.9$\sigma$ 
discrepancy between the third (2a) data points of simulations $A_{20}$ and $A_{30}$ on \fig\ref{fig:ratio}. 
If present, this effect comes
from $F_{\rm 2a}$, because the quantity $r_0^2F_1(r)$ is perfectly consistent between 
$A_{20}$ and $A_{30}$ (see \tab\ref{tab:Finterpol}).

It is  clear from \fig\ref{fig:ratio}  that 
the (absolute) discretization errors are bounded by 0.001.
The force ratio $F_{\rm 2a}/F_1$ at distance $xr_0$ 
has been interpolated to a few values of $x$ in \tab\ref{tab:F2a-interpol}.
The two-parameter function $f(x) = px^2/(x^2+q^2)$ was used to interpolate
the quantity plotted on \fig\ref{fig:ratio}, but the result never actually differs 
by more than one standard deviation from a simple linear interpolation in $x$.
\tab\ref{tab:F2a-interpol} also gives the result of 
a continuum extrapolation performed along the same lines as for $r_0^2F_1(r)$.

\begin{figure}
\begin{center}
\psfig{file=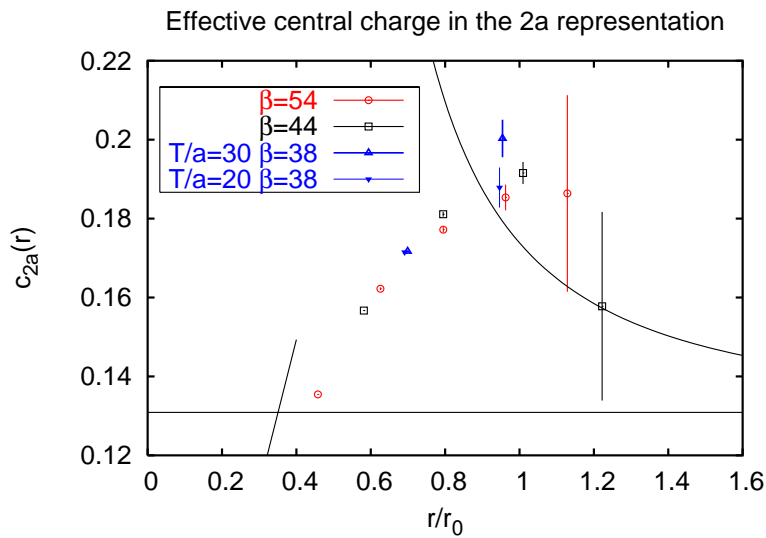,angle=0,width=10.5cm}
\end{center}
\vspace{-0.3cm}
\caption{The SU(5) effective central charge in the 2a representation.
The perturbative prediction is shown at short distance and the Nambu-Goto prediction at long distance.
The horizontal line marks the value $\pitf$.}
\la{fig:ceff2a}
\end{figure}

Since there is a stable string in the $\N$-ality 2 sector, it makes sense 
to consider the effective central charge in the 2a representation;
\fig\ref{fig:ceff2a} shows our data on $c_{\rm 2a}(r)$.
At short distances, it is roughly a factor 3/2 larger than in the fundamental 
representation, as predicted by perturbation theory. The curve seems to flatten off,
as it does in the fundamental representation, but the data does not extend
far enough to show whether it will bend down towards $\pitf$,
as we expect it to~\cite{meyer-teper-largeNconfinement}.
It was observed~\cite{altes-thick} in closed $k=2$
SU(8) strings that $c_{\rm 2a}<\frac{\sigma_{\rm 2a}}{\sigma_1}c_1$.

\subsection{Mixed representation correlators}
It is interesting to look at the cross correlation $\<P_{\rm 2a}(x)^* P_{\rm 2s}(x+r\hat1)\>$.
There is no symmetry which makes this correlator vanish. However we find that it is consistent
with zero. For instance, the `overlap'
\[
\frac{\<P_{\rm 2a}(x)^* P_{\rm 2s}(x+r\hat1)\>}{\sqrt{\Gamma_{\rm 2a}(r) \Gamma_{\rm 2s}(r)}}
\]
is consistent with zero in simulation $B$ with a statistical uncertainty of $5\cdot10^{-5}$,
$3\cdot10^{-4}$ and $4\cdot10^{-3}$ at distances $r=2a,~3a$ and $4a$ respectively.
Since
\[
\<P_{\rm 2a}(x)^* P_{\rm 2s}(y)\> =
{\textstyle\frac{1}{4}}
\left(\<(\tr\{U_P^*\})^2(x)~(\tr\{U_P\})^2(y) \>   
- \<\tr\{U_P^2\}^*(x)~\tr\{U_P^2\}(y) \>  \right),
\]
the statement that the cross correlation vanishes is equivalent to the property
that the two correlators on the RHS of this equation  are equal.

The interpretation that we propose is based on the observation that the direct product of 
the representations $({\rm 2a})^*$ and 2s does not contain the 
singlet representation (though it does contain the adjoint representation). 
Therefore if we imagine inserting 
the static particles adiabatically into the system,
they initially have an overall color charge. 
The only way the free energy can be made finite is if virtual adjoint charges 
screen the system. Since the external color charge object is extended,
it presumably costs a large amount of energy to screen it. 
In addition, statistically speaking the color of the virtual adjoint charge
must match that of the static system, which happens with a probability of order $1/N^2$.
Thus one expects the overlap to be $1/N^2$ color-phase-space suppressed~\cite{teper-lucini}.
All this results in a very small cross-correlation of the Polyakov loops.
This  large-$N$ argument was already made in~\cite{altes-thick} to explain
the fact that in closed $k$-string computations~\cite{teper-lucini,altes-thick},
the wavefunction of the lightest state is very well approximated by 
a (fuzzy) Polyakov loop in the totally antisymmetric representation of 
$\N$-ality $k$.

\newpage
\section{Conclusion\label{sec:conclu}}
We computed  the static force in the $d=3$ SU(5) gauge theory employing an 
efficient three-level algorithm. Linear confinement is observed, and
the effective string tension defined in \eq\ref{eq:sigeff} is essentially
constant beyond 0.5fm. The string tension extracted at 1fm agrees
with the closed string tension extracted from torelon spectroscopy~\cite{teper2+1}.

By comparison with existing SU(2)~\cite{pushan,caselle} and SU(3)~\cite{lw-bosonic} data, 
the quantity $\sigma r_0^2$ is found to be  independent of the 
number of colors at the $1\%$ level; it also has very small discretization errors.
Thus comparing SU($N$) gauge theories at common string tension or
common Sommer reference scale is equivalent at that level of precision.

The effective central charge $c(r)$ was obtained in the range 0.25---0.75fm. It 
converges to within $5\%$ of the expected asymptotic value of $\pitf$, 
confirming the multiplicity and the bosonic nature of the flux-tube's massless degrees
of freedom. A comparison with the SU(2) and SU(3) data reveals that the distance
$\rs$ where $c(r)$ crosses the value $\pitf$ decreases steadily by almost a factor two
from SU(2) to SU(5). Since (bosonic) effective string theory predicts that 
the asymptotic value is approached from above~\cite{lw-beyond}, it is tempting to speculate
that the asymptotic expansion's accuracy is higher at a fixed value of $\sqrt{\sigma}r$
for larger $N$. 

We also studied the static force in the symmetrized (2s) and antisymmetrized (2a)
direct product representations of two `quarks'. For $N\geq4$,
it is known~\cite{teper-lucini} that such 
static charges of $\N$-ality 2 are linearly confined with 
a different string tension from the fundamental one. And indeed
we find that the ratio $F_{\rm 2a}/F_1$ is constant to a first approximation, 
confirming the linear confinement property. Furthermore, it is very 
close to the Casimir scaling prediction (3/2 in this case).
A more detailed study reveals that deviations are present at the $0.1\%$
to $1\%$ level in the range 0.2---0.7fm. They are positive, unlike
the case of the adjoint static force~\cite{krato} where screening of the adjoint 
flux must set in at some distance $r$. The 2s representation on the other hand
exhibits a negative deviation, as one expects 
if there is a single stable string per $\N$-ality and screening occurs.

Since the representation 2a has $\N$-ality 2 and has the smallest quadratic Casimir
of all irreducible representations in that sector, one might expect to find
again a precocious onset of the effective central charge $c_{\rm 2a}(r)$
for the $k=2$ string. This is not the case: at $r=r_0$, where 
$c(r)$ lies within a few percent  of $\pitf$, $c_{\rm 2a}(r)$ is almost as large
as $\frac{3}{2}\cdot \pitf$. Data at further distances is needed to see
whether it decreases towards $\pitf$. 

\vspace{1cm}

\subsection*{Acknowledgements\label{sec:acknow}}
I thank the computer team of DESY/NIC for providing an efficient
batch system. I am indebted to Pushan Majumdar for communicating unpublished
SU(2) data and reading the manuscript. 
I also thank York Schr\"oder for discussions and
Rainer Sommer for a critical reading of the manuscript.

\appendix\subsection*{Appendix: discretization errors and the continuum limit}
The quantity $r_0^2F(r)$ has been extrapolated linearly in $(a/r_0)^2$
to the continuum, using 
the data from simulations $B$ and $C$. This is our best estimate in the continuum 
limit and it can be found in \tab\ref{tab:Finterpol}. The first number in brackets is the 
statistical error on the result (there is no $\chi^2$). To estimate the 
systematic uncertainty on this procedure, we also do a quadratic fit in $(a/r_0)^2$
to simulations $A_{30}$, $B$ and $C$ (again, there is no $\chi^2$). 
The second number given in brackets
is the continuum value obtained by quadratic extrapolation minus
the quoted continuum result. Thus a conservative error estimate is the maximum
of the absolute value of the two numbers in brackets.
The continuum limit is illustrated for $r_0^2\sigma(r)$, 
which is equivalent to $r_0^2F(r)$ for this purpose, on \fig\ref{fig:conti}.

We proceeded in the same way for $F_{2a}(r)/F_1(r)$ (\tab\ref{tab:F2a-interpol}).
For the effective central charge, we also give the continuum $c(r)$
obtained in this way in \tab\ref{tab:cinterpol}. Recall that the definition of Ref.~\cite{lw-bosonic} 
was used to define $c(r)$ at finite lattice spacing. We now show 
that a different definition gives the same result in the continuum limit.

We propose to determine the parameters
$a^2\tilde \sigma(r),~ a\tilde \mu(r)$ and $\tilde c(r)$ by 
\ba
\tilde c(r)      &=& -\frac{r(r-a)(r+a)}{2a^2}~\left(V(r-a)-2V(r)+V(r+a)\right) \\
a^2\tilde \sigma(r) &=& \half(r-a)~V(r-a) - r~ V(r) + \half(r+a)~V(r+a)  \\
a\tilde \mu(r)    &=&  -\frac{(r-a)(r+\frac{a}{2})}{a} ~V(r-a) + \frac{2r^2}{a} ~V(r)
                       -\frac{(r+a)(r-\frac{a}{2})}{a}~ V(r+a).
\ea
This new definition is such that if the on-axis lattice Polyakov loop correlator 
was given by 
\be
e^{-V_0(r)T},\qquad V_0(r) = \mu +\sigma r - \frac{c}{r},
\ee
then the effective parameters match those appearing in $V_0(r)$ exactly.
Note that the functions $c$ and $\tilde c$ differ only by O($a^2$) terms, 
and that $\tilde c(r)$ is simply a factor $(1-(a/r)^2)$ times the `naive'
definition of the effective central charge used in~\cite{pushan}.
And indeed, comparing \tab\ref{tab:ctilinterpol} to
\tab\ref{tab:cinterpol}, we find that 
$\tilde c(r)$ is compatible with $c(r)$ in the continuum limit, 
as it should be; see \fig\ref{fig:conti} for an illustration. 
This consistency-check is also a test of our error analysis.

\begin{figure}
\begin{center}
\psfig{file=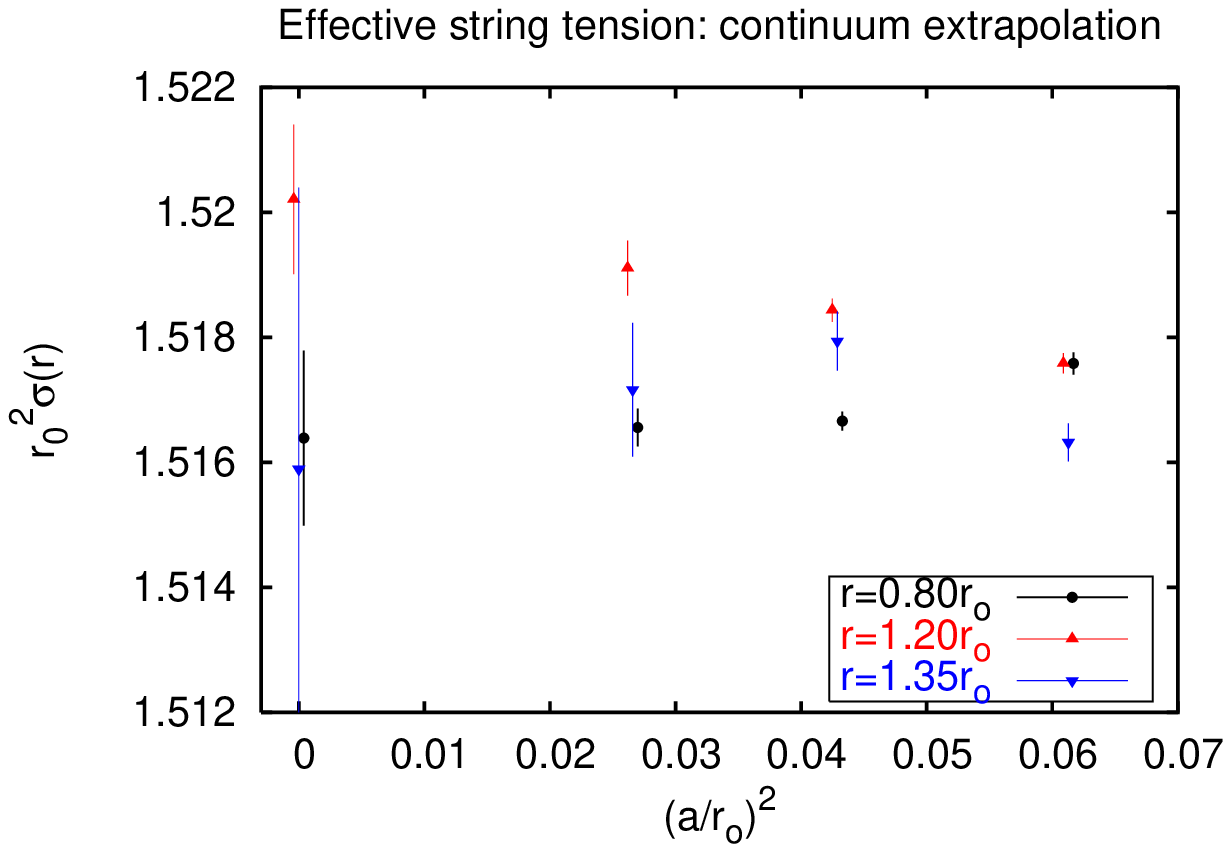,angle=0,width=12.5cm}\\
\vspace{1cm}
\psfig{file=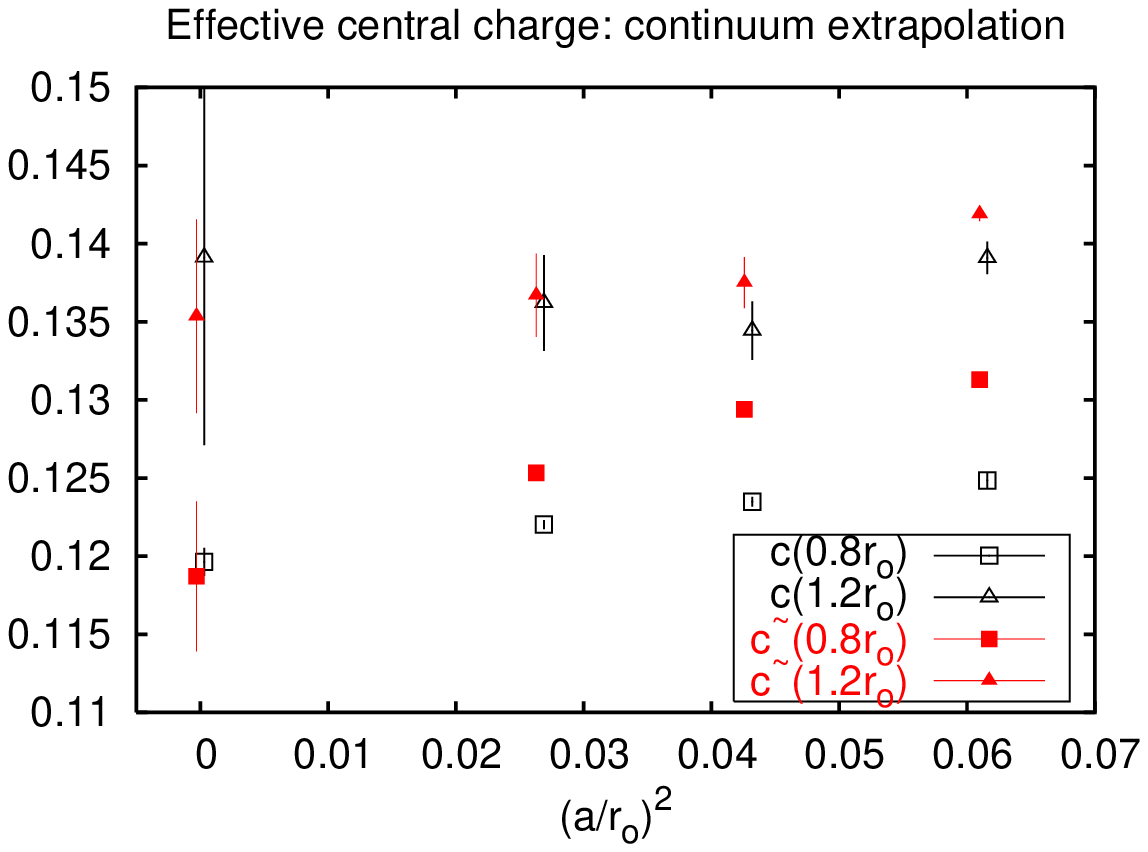,angle=0,width=12.5cm}
\end{center}
\caption{Continuum extrapolation of the effective string tension 
$r_0^2\sigma(r)=r_0^2 F(r)-\pitf (r_0/r)^2$
and two definitions of the central charge 
$c(r)$ and $\tilde c(r)$ in $d=3$ SU(5) gauge theory.
The result, with its final error bar, is shown at vanishing abscissa.}
\la{fig:conti}
\end{figure}


\newpage
\appendix

\begin{table}
\begin{center}
\begin{tabular}{c@{\qquad}l@{\qquad}l@{\qquad}l}
\hline\hline
$r/a$  & $\Gamma_1(r)$   & $\Gamma_{2a}(r)$   &$\Gamma_{2s}(r)$    \\
\hline
2 & $  7.4136(20)\cdot10^{-4} $ &$ 2.00130(80)\cdot10^{-5} $ & $ 5.1040(32)\cdot10^{-8} $  \\
3 & $  8.6971(40)\cdot10^{-5} $ &$ 8.0083(57)\cdot10^{-7} $ & $ 3.5219(40)\cdot10^{-10} $  \\
4 & $  1.29486(95)\cdot10^{-5}$ &$ 4.5651(52)\cdot10^{-8} $ & $ 4.332(24)\cdot10^{-12}$   \\
5 & $  2.1498(22)\cdot10^{-6} $ &$ 3.0544(48)\cdot10^{-8} $ & $ 5.41(93)\cdot10^{-14}  $  \\
6 & $  3.7784(53)\cdot10^{-7} $ &$ 2.2198(54)\cdot10^{-10} $ &  --    \\
7 & $  6.862(12)\cdot10^{-8}  $ &$ 1.676(10)\cdot10^{-11} $ &   --  \\
8 & $  1.2705(31)\cdot10^{-8} $ &$ 1.317(74)\cdot10^{-12}  $ &   --   \\
9 & $  2.3909(81)\cdot10^{-9} $ &$ 1.36(38)\cdot10^{-13}  $ &  --    \\
10& $  4.553(37)\cdot10^{-10} $ & -- &  --  \\
\hline\hline
\end{tabular}
\end{center}
\caption{The raw data of simulation B.}
\la{tab:PP44}
\end{table}

\begin{table}
\begin{center}
\begin{tabular}{c@{\qquad}l@{\qquad}l@{\qquad}l}
\hline\hline
$\bar r/a$ & $a^2F_1(\bar r)$ & $F_{2a}(\bar r)/F_1(\bar r)$ &$F_{2s}(\bar r)/F_1(\bar r)$  \\
\hline
2.379 &  0.0857165(97) &    1.501928(46)&    2.32216(20)   \\
3.407 &  0.076183(12)  &    1.504059(99)&    2.3092(28)   \\
4.432 &  0.071825(15)  &    1.50613(20) &    2.424(95)   \\
5.448 &  0.069546(18)  &    1.50793(71) &    --         \\
6.458 &  0.068237(23)  &    1.5144(33)  &    --        \\
7.464 &  0.067463(39)  &    1.507(33)   &    --       \\
8.469 &  0.066812(67)  &    --          &    --      \\
9.473 &  0.06634(25)   &    --          &    --     \\
\hline\hline
\end{tabular}
\end{center}
\caption{The static forces in simulation B.}
\la{tab:Fr}
\end{table}

\begin{table}
\begin{center}
\begin{tabular}{c@{\qquad}l@{\qquad}l@{\qquad}l}
\hline\hline
$\tilde r/a$ & $c_1(\tilde r)$ & $c_{2a}(\tilde r)$ & $c_{2s}(\tilde r)$ \\
\hline
2.808 &  0.105535(54)  &  0.15671(10) &   0.2560(23) \\
3.838 &  0.12320(31)   &  0.18110(61) & --          \\
4.875 &  0.13202(48)   &  0.1916(28)  & --         \\
5.902 &  0.1346(21)    &  0.158(24)   & --        \\ 
6.920 &  0.1283(45)    &  --          & --       \\
7.932 &  0.162(16)     &  --          & --      \\
\hline\hline
\end{tabular}
\end{center}
\caption{The effective central charges in simulation B.}
\la{tab:cr}
\end{table}

\begin{table}
\begin{center}
\begin{tabular}{c@{\quad}l@{\qquad}l@{\qquad}l@{\qquad}l@{\qquad}l}
\hline\hline
$ r/r_0$ & $A_{20}$ & $A_{30}$ & $B$ & $C$ & cont.\\
\hline

{\sm 0.50} &{\sm   --         }  &{\sm   --         } &{\sm  1.98604(28) }&{\sm  1.97861(46)} &{\sm  ---  } \\ 
{\sm 0.65} &{\sm   1.81564(22)}  &{\sm   1.81552(18)}  &{\sm  1.81409(28)}  &{\sm  1.81380(35)} &{\sm  1.8133(10)(20)}\\
{\sm 0.80} &{\sm   1.72229(22)}  &{\sm   1.72211(18)}  &{\sm  1.72119(15)}  &{\sm  1.72108(30)} &{\sm  1.7209(8)(14)} \\
{\sm 0.95} &{\sm   1.664406(60)} &{\sm  1.664354(48)}  &{\sm  1.664300(52)} &{\sm  1.66417(12)} &{\sm  1.6640(3)(-2) }\\
{\sm 1.00} &{\sm   1.65       }  &{\sm   1.65       }  &{\sm    1.65    }   &{\sm  1.65      }  &{\sm  1.65 }\\
{\sm 1.05} &{\sm   1.637603(52)} &{\sm  1.637647(41)}  &{\sm  1.637694(45)} &{\sm  1.63780(10)} &{\sm  1.6380(3)(1)} \\
{\sm 1.20} &{\sm   1.60865(24)}  &{\sm   1.60849(16)}  &{\sm  1.60934(19)}  &{\sm  1.61001(44)} &{\sm  1.6111(12)(-2)}\\
{\sm 1.35 }&{\sm   1.58855(46)}  &{\sm    1.58815(30)} &{\sm 1.58977(47)}  &{\sm  1.5890(11) } &{\sm  1.5877(29)(-45)}\\
{\sm 1.50} &{\sm   1.57272(81)}  &{\sm   1.57386(65)}  &{\sm  1.57638(76)}  &{\sm  1.5686(32)}  &{\sm  1.556(9)(-20)}\\
{\sm 1.65} &{\sm   1.5615(18) }  &{\sm    1.5635(19)}  &{\sm  1.5647(10) }  &{\sm    --     }   & {\sm -- } \\
{\sm 1.80} &{\sm   1.5536(47) }  &{\sm  1.5558(55) }   &{\sm  1.5551(23) }  &{\sm  --       }   &{\sm  -- } \\
{\sm 2.00} &{\sm   1.562(13)  }  &{\sm   1.548(32) }   &{\sm  1.5451(65) }  &{\sm  --       }   &{\sm  -- } \\
\hline\hline
\end{tabular}
\end{center}
\caption{The quantity $r_0^2F(r)$ interpolated, and its continuum limit (details in the App.).}
\la{tab:Finterpol}
\end{table}

\begin{table}
\begin{center}
\begin{tabular}{c@{\quad}l@{\qquad}l@{\qquad}l@{\qquad}l@{\qquad}l}
\hline\hline
$r/r_0$ & $A_{20}$ & $A_{30}$ & $B$ & $C$ & cont.\\
\hline
{\sm0.458} & {\sm  --         }  & {\sm         -- }  & {\sm     --    } & {\sm0.090714(40)}& {\sm--}\\ 
{\sm0.65 }& {\sm  0.11028(74) } & {\sm 0.11141(56) } & {\sm  0.11284(27)} &{\sm 0.11163(19)} &{\sm 0.1096(7)(-50)} \\  
{\sm0.80} & {\sm  0.12573(64)}  & {\sm 0.12484(50) } & {\sm  0.12348(31)} &{\sm 0.12202(28)} &{\sm 0.1196(9)(-5)} \\
{\sm0.95 }& {\sm  0.13282(56)}  & {\sm 0.13235(44) } & {\sm  0.13006(36)} &{\sm 0.1304(10) } &{\sm 0.131(3)(5)} \\
{\sm1.00} & {\sm  0.13424(57)}  &{\sm  0.13412(42) } &{\sm   0.13172(46)} &{\sm 0.1314(11)}  &{\sm 0.131(3)(4)} \\
{\sm1.05} & {\sm  0.13534(71)}  &{\sm  0.13564(48) } &{\sm   0.13279(44)} & {\sm0.1315(12)}  &{\sm 0.129(3)(2)} \\
{\sm1.20} & {\sm  0.1374(15) }  & {\sm 0.1390(10)  } &{\sm   0.1344(19) } & {\sm0.1362(31)}  &{\sm 0.139(8)(12)}\\
\hline\hline
\end{tabular}
\end{center}
\caption{The effective central charge $c(r)$ interpolated, and its continuum limit.}
\la{tab:cinterpol}
\end{table}

\begin{table}
\begin{center}
\begin{tabular}{c@{\quad}l@{\qquad}l@{\qquad}l@{\qquad}l@{\qquad}l}
\hline\hline
$r/r_0$ & $A_{20}$ & $A_{30}$ & $B$ & $C$ & cont.\\
\hline
{\sm0.458}& {\sm  --         }  & {\sm         -- }  & {\sm     --    }   & {\sm0.098331(44)}& {\sm--}\\ 
{\sm0.65 }& {\sm  0.1105(19) } & {\sm 0.1135(15) }  & {\sm  0.11787(15)} &{\sm 0.11611(21)} &{\sm 0.1132(6)(-114)} \\  
{\sm0.80} & {\sm  0.13196(48)}  & {\sm 0.13130(37) } & {\sm  0.12939(35)} &{\sm 0.12533(27)} &{\sm 0.1187(9)(-48) } \\
{\sm0.95 }& {\sm  0.13995(64)}  & {\sm 0.13921(51) } & {\sm  0.13491(33)} &{\sm 0.13260(92)} &{\sm0.129(2)(3) } \\
{\sm1.00} & {\sm  0.14117(58)}  &{\sm  0.14074(46) } &{\sm   0.13606(42)} &{\sm 0.1340(11)}  &{\sm0.131(3)(4) } \\
{\sm1.05} & {\sm  0.14194(62)}  &{\sm  0.14190(45) } &{\sm   0.13690(45)} & {\sm0.1339(12)}  &{\sm 0.1289(32)(7)} \\
{\sm1.20} & {\sm  0.1425(14) }  & {\sm 0.14393(92) } &{\sm   0.1375(16) } & {\sm0.1367(27)}  &0.135(7)(6)\\ 
\hline\hline
\end{tabular}
\end{center}
\caption{The effective central charge $\tilde c(r)$ interpolated, and its continuum limit.}
\la{tab:ctilinterpol}
\end{table}

\begin{table}
\begin{center}
\begin{tabular}{c@{\qquad}l@{\qquad}l@{\qquad}l@{\qquad}l}
\hline\hline
$r/r_0$ &   $B$       &      $C$ \\
\hline
0.458 &      --      & 0.135447(60) \\  
0.65 &  0.16725(88)   & 0.16486(32) \\
0.80 &  0.18146(60)   & 0.17752(69) \\
0.95 &  0.1894(20)    & 0.1849(30) \\
1.00 &  0.1912(27)    & 0.1860(54) \\
1.05 &  0.1868(38)    & 0.186(12) \\
1.20 &  0.162(21)     & 0.135(54) \\
\hline\hline
\end{tabular}
\end{center}
\caption{The quantity $c_{\rm 2a}(r)$ interpolated.}
\la{tab:cinterpol2a}
\end{table}

\begin{table}
\begin{center}
\begin{tabular}{c@{\quad}l@{\qquad}l@{\qquad}l@{\qquad}l@{\qquad}l}
\hline\hline
$r/r_0$ & $A_{20}$ & $A_{30}$ & $B$ & $C$ & cont. \\
\hline
{\sm0.50 }&{\sm --    }      &{\sm --        } & {\sm 1.324(31)  }  &{\sm 1.130(39) }   &{\sm --  }\\
{\sm0.65 } & {\sm2.434(65) }  &{\sm 2.567(50) } & {\sm 2.279(50) }   &{\sm 2.087(48)  } &{\sm 1.77(15)(13) } \\ 
{\sm0.80 } & {\sm3.59(15) }  &{\sm 3.693(97) }  & {\sm 3.317(66) }   &{\sm 3.32(12) }  & {\sm3.3(3)(7)  }\\
{\sm0.95 } &{\sm 5.09(36) }  &{\sm 4.01(61)  }  & {\sm 4.28(15)  }   &{\sm 4.80(21)  }  &{\sm 5.7(6)(6)  }  \\
{\sm1.00 } &{\sm 5.66(67) }  &{\sm 3.94(81)  }  & {\sm 4.57(21)  }   &{\sm  5.35(29)  } & {\sm6.6(8)(4) } \\
{\sm1.05 } &{\sm 6.25(84)  } &{\sm 3.87(97)  }  & {\sm 4.86(30)  }   &{\sm 5.92(43) }  &{\sm -- } \\
{\sm1.20 } &{\sm --  } & {\sm8.5(3.2) }  & {\sm 6.57(59) }    & {\sm 7.0(2.2)  }  & {\sm -- }    \\
\hline\hline
\end{tabular}
\end{center}
\caption{The quantity $10^3\cdot(\frac{2}{3} F_{2a}(r)/F_1(r)-1)$ interpolated, and its continuum limit.}
\la{tab:F2a-interpol}
\end{table}

\begin{table}
\begin{center}
\begin{tabular}{c@{\qquad}l@{\qquad}l@{\qquad}l@{\qquad}l}
\hline\hline
$r/r_0$ &   $B$       &      $C$ \\
\hline
0.50 & -4.943(88)   & -4.55(10) \\
0.65 & -8.68(69)    & -8.73(57) \\
\hline\hline
\end{tabular}
\end{center}
\caption{The quantity $10^3\cdot(\frac{3}{7} F_{2s}(r)/F_1(r)-1)$ interpolated.}
\la{tab:F2s-interpol}
\end{table}


\newpage
\begin{thebibliography}{99}
\bibitem{ALPHA}
M.~L\"uscher, R.~Sommer, P.~Weisz and U.~Wolff,
Nucl.\ Phys.\ B {\bf 413} (1994) 481
[arXiv:hep-lat/9309005].
\bibitem{polstrom}
J.~Polchinski and A.~Strominger,
Phys.\ Rev.\ Lett.\  {\bf 67} (1991) 1681.
\bibitem{lw-beyond}
M.~L\"uscher and P.~Weisz,
JHEP {\bf 0407} (2004) 014
[arXiv:hep-th/0406205].
\bibitem{teper2+1}
M.~J.~Teper,
Phys.\ Rev.\ D {\bf 59}, 014512 (1999)
[arXiv:hep-lat/9804008].
\bibitem{mikko}
M.~Laine,
arXiv:hep-ph/0301011.
\bibitem{hooft-largeN}
G.~'t Hooft,
Nucl.\ Phys.\ B {\bf 72} (1974) 461.
\bibitem{regge2+1}
H.~B.~Meyer and M.~J.~Teper,
Nucl.\ Phys.\ B {\bf 668}, 111 (2003)
[arXiv:hep-lat/0306019].
\bibitem{thesis}
H.~B.~Meyer,
arXiv:hep-lat/0508002.
\bibitem{luscher81}
M.~L\"uscher, K.~Symanzik and P.~Weisz,
Nucl.\ Phys.\ B {\bf 173} (1980) 365;
M.~L\"uscher,
Nucl.\ Phys.\ B {\bf 180} (1981) 317.
\bibitem{lw-bosonic}
M.~L\"uscher and P.~Weisz,
JHEP {\bf 0207} (2002) 049
[arXiv:hep-lat/0207003].
\bibitem{HariDass:2005we}
N.~D.~Hari Dass and P.~Majumdar,
PoS {\bf LAT2005} (2006) 312
[arXiv:hep-lat/0511055].
\bibitem{dalley}
S.~Dalley,
arXiv:hep-th/0512264.
\bibitem{kuti_dublin}
J.~Kuti,
PoS {\bf LAT2005} (2005) 001
[arXiv:hep-lat/0511023].
\bibitem{lw-algo}
M.~L\"uscher and P.~Weisz,
JHEP {\bf 0109} (2001) 010
[arXiv:hep-lat/0108014].
\bibitem{sommer}
R.~Sommer,
Nucl.\ Phys.\ B {\bf 411} (1994) 839
[arXiv:hep-lat/9310022].
\bibitem{ambjorn3d}
J.~Ambj{\o}rn, P.~Olesen and C.~Peterson,
Nucl.\ Phys.\ B {\bf 240} (1984) 533.
\bibitem{ambjorn4d}
J.~Ambj{\o}rn, P.~Olesen and C.~Peterson,
Nucl.\ Phys.\ B {\bf 240} (1984) 189.
\bibitem{greensite}
M.~Faber, J.~Greensite and S.~Olejnik,
Phys.\ Rev.\ D {\bf 57} (1998) 2603
[arXiv:hep-lat/9710039].
\bibitem{cornwall}
J.~M.~Cornwall,
Phys.\ Rev.\ D {\bf 57} (1998) 7589
[arXiv:hep-th/9712248].
\bibitem{simonov}
Y.~A.~Simonov,
JETP Lett.\  {\bf 71} (2000) 127
[arXiv:hep-ph/0001244];
V.~I.~Shevchenko and Y.~A.~Simonov,
Phys.\ Rev.\ Lett.\  {\bf 85} (2000) 1811
[arXiv:hep-ph/0001299].
\bibitem{deldar}
S.~Deldar,
Phys.\ Rev.\ D {\bf 62} (2000) 034509
[arXiv:hep-lat/9911008].
\bibitem{bali}
G.~S.~Bali,
Phys.\ Rev.\ D {\bf 62} (2000) 114503
[arXiv:hep-lat/0006022].
\bibitem{piccioni}
C.~Piccioni,
Phys.\ Rev.\ D {\bf 73} (2006) 114509
[arXiv:hep-lat/0503021].
\bibitem{poulis}
G.~I.~Poulis and H.~D.~Trottier,
Phys.\ Lett.\ B {\bf 400} (1997) 358
[arXiv:hep-lat/9504015].
\bibitem{krato}
S.~Kratochvila and P.~de Forcrand,
Nucl.\ Phys.\ B {\bf 671} (2003) 103
[arXiv:hep-lat/0306011].
\bibitem{Stephenson:1999kh}
P.~W.~Stephenson,
Nucl.\ Phys.\ B {\bf 550} (1999) 427
[arXiv:hep-lat/9902002].
\bibitem{Philipsen:1999wf}
O.~Philipsen and H.~Wittig,
Phys.\ Lett.\ B {\bf 451} (1999) 146
[arXiv:hep-lat/9902003].
\bibitem{teper-lucini}
B.~Lucini and M.~Teper,
Phys.\ Rev.\ D {\bf 64}, 105019 (2001)
[arXiv:hep-lat/0107007].
\bibitem{schroeder}
Y.~Schr\"oder,
Phys.\ Lett.\ B {\bf 447} (1999) 321
[arXiv:hep-ph/9812205];
Y.~Schr\"oder,
DESY-THESIS-1999-021
\bibitem{altes-thick}
C.~P.~Korthals Altes and H.~B.~Meyer,
arXiv:hep-ph/0509018.
\bibitem{pushan}
P.~Majumdar,
arXiv:hep-lat/0406037.
\bibitem{pushan-pc} H.~D.~Hari Dass, P.~Majumdar, in preparation.
\bibitem{caselle}
M.~Caselle, M.~Pepe and A.~Rago,
JHEP {\bf 0410} (2004) 005
[arXiv:hep-lat/0406008].
\bibitem{kari}
M.~Laine, H.~B.~Meyer, K.~Rummukainen and M.~Shaposhnikov,
JHEP {\bf 0404}, 027 (2004)
[arXiv:hep-ph/0404058].
\bibitem{martinh}
M.~Hasenbusch and S.~Necco,
JHEP {\bf 0408} (2004) 005
[arXiv:hep-lat/0405012].
\bibitem{inprep} H.~B.~Meyer, in preparation.
\bibitem{multihit}
G.~Parisi, R.~Petronzio and F.~Rapuano,
Phys.\ Lett.\ B {\bf 128} (1983) 418.
\bibitem{Cabibbo:1982zn}
N.~Cabibbo and E.~Marinari,
Phys.\ Lett.\ B {\bf 119} (1982) 387.
\bibitem{fabhaan}
K.~Fabricius and O.~Haan,
Phys.\ Lett.\ B {\bf 143} (1984) 459.
\bibitem{kenpen}
A.~D.~Kennedy and B.~J.~Pendleton,
Phys.\ Lett.\ B {\bf 156} (1985) 393.
\bibitem{adler-or}
S.~L.~Adler,
Phys.\ Rev.\ D {\bf 23} (1981) 2901.
\bibitem{uwerr}
U.~Wolff  [ALPHA collaboration],
Comput.\ Phys.\ Commun.\  {\bf 156} (2004) 143
[arXiv:hep-lat/0306017].
\bibitem{sommer-necco}
S.~Necco and R.~Sommer,
Nucl.\ Phys.\ B {\bf 622} (2002) 328
[arXiv:hep-lat/0108008].
\bibitem{arvis}
J.~F.~Arvis,
Phys.\ Lett.\ B {\bf 127} (1983) 106.
\bibitem{meyer-teper-largeNconfinement}
H.~Meyer and M.~Teper,
JHEP {\bf 0412}, 031 (2004)
[arXiv:hep-lat/0411039].
\end{thebibliography}
\end{document}